\documentclass[a4,11pt]{article}
\usepackage{color}
\usepackage{bm}
\usepackage{slashed}
\usepackage{xcolor} 
\usepackage{graphicx}
\usepackage{amsmath}
\usepackage{amssymb}
\usepackage{amsfonts}
\usepackage{amsthm}
\usepackage{array}

\usepackage[
      colorlinks=true,
      linkcolor=blue,
      urlcolor=blue,
      filecolor=blue,
      citecolor=red,
      pdfstartview=FitV,
      pdftitle={},
      pdfauthor={},
      pdfsubject={},
      pdfkeywords={},
      pdfpagemode=None,
      bookmarksopen=true
]{hyperref}

\usepackage{epsfig}

\usepackage{hyperref}
\def\beq{\begin{eqnarray}}
\def\eeq{\end{eqnarray}}

\def\be{\begin{equation}}
\def\ee{\end{equation}}
\def\bea{\begin{eqnarray}}
\def\eea{\end{eqnarray}}

\def\be{\begin{equation}}
\def\ee{\end{equation}}
\def\bea{\begin{eqnarray}}
\def\eea{\end{eqnarray}}

\def\nn{\nonumber}

\newcounter{mnotecount}[section]

\renewcommand{\themnotecount}{\thesection.\arabic{mnotecount}}

\newcommand{\mnotex}[1]
{\protect{\stepcounter{mnotecount}}$^{\mbox{\footnotesize
$
\bullet$\themnotecount}}$ \marginpar{
\raggedright\tiny\em
$\!\!\!\!\!\!\,\bullet$\themnotecount: #1} }

\numberwithin{equation}{section}

\usepackage{mathrsfs}

\begin{document}

 \begin{centering}
  
{\Large  {\bf
On supertranslation invariant Lorentz charges } \\

 \vspace{0.8cm}
{\normalsize Sumanta Chakraborty$^{a}$,
Sk Jahanur Hoque$^{b,c,d}$,  Amitabh Virmani$^{e}$
\vspace{.8cm} \\
\begin{minipage}{.8\textwidth}
\small  
\begin{center}
$^{a}${\small{School of Physical Sciences, Indian Association for the Cultivation of Science, \\ Kolkata 70032, India}}
\vspace{0.3cm} \\
$^{b}${\small{Birla Institute of Technology and Science, Pilani, Hyderabad Campus,}}
\\
{\small{Jawaharnagar, Hyderabad 500 078, India}}
\vspace{0.3cm} \\
$^{c}${\small{Institute of Theoretical Physics,
Faculty of Mathematics and Physics, Charles University,}}
\\ 
{\small{V~Hole\v{s}ovi\v{c}k\'ach 2, 180~00 Prague 8, Czech Republic}}
\vspace{0.3cm} \\
$^{d}${\small{Universit\'{e} Libre de Bruxelles, International Solvay Institutes,}}
\\
{\small{CP 231, B-1050 Brussels, Belgium}}
\vspace{0.3cm} \\
$^{e}${\small{Chennai Mathematical Institute, H1, SIPCOT IT Park, Siruseri,}}
\\
{\small{Kelambakkam 603103, Tamil Nadu, India}}
\end{center}
\end{minipage}
 \vspace{0.8cm}

{\tt tpsc@iacs.res.in, jahanur.hoque@hyderabad.bits-pilani.ac.in, \\ avirmani@cmi.ac.in}

}

}
\end{centering}

\begin{abstract}
In recent papers, Fuentealba, Henneaux, and Troessaert (FHT) gave definitions for supertranslation invariant Lorentz charges in the ADM Hamiltonian formalism and showed that their definitions match with the Chen, Wang, Yau (CWY) definitions of  Lorentz charges at null infinity which are free from ``supertranslation ambiguities''. In this brief note, motivated by the analysis of FHT, we write expressions for the supertranslation invariant Lorentz charges in Beig-Schmidt variables at spacelike and timelike infinity. We present calculations, building upon the work of Comp\`ere, Gralla, and Wei (CGW), to show that our expressions for supertranslation invariant Lorentz charges match the CWY definitions at null infinity.
\end{abstract}

\newpage
\tableofcontents
\section{Introduction and summary}

Working in the ADM Hamiltonian framework, Fuentealba, Henneaux, and Troessaert (FHT) in a recent series of papers \cite{ Fuentealba:2022xsz, Fuentealba:2023syb} (and also \cite{Henneaux:2023neb}) have shown that the symmetries of the asymptotically flat spacetimes at spacelike infinity can be extended from Bondi-Metzner-Sachs (BMS) symmetries \cite{Bondi et al.(1962),Sachs(1962)} (see \cite{Henneaux:2019yax} for a review) to include logarithmic supertranslations. At the present state of development, it is not understood how these logarithmic supertranslations symmetries act on Bondi variables at null infinity or on Beig-Schmidt variables at spacelike or timelike infinity.

A key feature of the logarithmic supertranslations, as they appear in the work of FHT, is that the charges generating logarithmic supertranslations are canonically conjugate to the charges generating the standard supertranslations. This feature allowed FHT to define Lorentz charges such that they have vanishing commutators with the standard supertranslations charges. These charges are called ``supertranslation invariant Lorentz charges''.

In completely independent developments, mostly predating the work of FHT, certain redefinitions of Lorentz charges were put forward which are free from ``supertranslation ambiguities'' at null infinity \cite{Chen:2021szm, Chen:2022fbu}.\footnote{For further references and discussions, see \cite{Chen:2022fbu, Compere:2021inq, Javadinezhad:2022ldc, Ashtekar:2023zul}.} Following \cite{Chen:2022fbu}, we call these Lorentz charges Chen, Wang, Yau (CWY) charges. In ref.~\cite{Fuentealba:2023syb}, FHT showed that the supertranslation invariant Lorentz charges they defined are the same as the CWY charges where the matching with null infinity was done by going to hyperbolic coordinates at spatial infinity.  As such, there is no parallel between the work of FHT and the works at null infinity. The reasoning with which  FHT  arrived at their expression is closely tied to the Hamiltonian framework and the extension of BMS symmetries to include logarithmic supertranslations. As mentioned in the opening paragraph, the analog of these logarithmic supertranslations at null infinity is presently not understood.

In recent years, the matching of various quantities from future to past null infinity via spacelike infinity has received much attention. This is natural from the point of view of an S-matrix approach where data on future null infinity are related to the ones on past null infinity. The matching is done through spacelike infinity~\cite{Strominger:2013jfa, Strominger:2017zoo}. Comp\`ere, Gralla, and Wei (CGW) \cite{Compere:2023qoa} presented a detailed study of matching various fields, charge expressions, transformation laws, etc., across various infinities.

In this brief note, motivated by the analysis of FHT, (i) we write expressions for supertranslation invariant Lorentz charges in Beig-Schmidt variables at spacelike and timelike infinity, (ii) we present calculations building upon the work of CGW to show that our definitions are the same as the CWY definitions of Lorentz charges at null infinity. 
To the best of our knowledge, such an analysis is missing in the literature, and since Beig-Schmidt variables  \cite{BeigSchmidt, Beig:1983sw}
 have been a useful source of insights, we hope that our analysis is of value. 
 
 Admittedly, our analysis is limited in scope. We do not perform a Hamiltonian analysis in Beig-Schmidt variables. In fact, detailed boundary conditions that allow for logarithmic supertranslations in the Beig-Schmidt variables are also not fully understood.\footnote{Some comments appear in \cite{Compere:2011ve, Compere:2023qoa}.} We hope to explore this subject further in our future work.

The rest of the note is organized as follows. In section \ref{sec:our_defs}, we present expressions for supertranslation invariant Lorentz charges in Beig-Schmidt variables at spacelike and timelike infinity. In section \ref{sec:matching_CWY}, we connect our invariant charges to null infinity. In section \ref{sec:CWY_and_others}, we compare different definitions of invariant charges at null infinity.

The idea of writing this note grew out of a discussion at the Quantum Gravity at the Raman Research Institute (QG at RRI) meeting in September 2023. At the meeting, we spoke on very different topics, but among ourselves, we discussed this topic the most. When the organizers contacted us to contribute an article to the special volume of Gen.~Rel.~Grav.~we all agreed that this would be the most fitting contribution.

\section{Supertranslation invariant Lorentz charges at spacelike/timelike infinity}
\label{sec:our_defs}
We start with a discussion of spacelike infinity; the adaptation to timelike infinity is straightforward \cite{Chakraborty:2021sbc, Compere:2023qoa}. For spacelike infinity, our starting point is the Beig-Schmidt ansatz (equation (188) of Ref.~\cite{Compere:2023qoa}),
\begin{align}
ds^2 &= \left(1+\frac{2\sigma}{\rho} +\frac{\sigma^2}{\rho^2} + o(\rho^{-2}) \right)d\rho^2 +  o(\rho^{-2})\rho d\rho d\phi^a
\nonumber
\\
& + \rho^2\biggr(h_{ab} + \rho^{-1}\big(k_{ab}-2\sigma h_{ab}\big) +\rho^{-2}\log\rho \, i_{ab}+ \rho^{-2}j_{ab} +o(\rho^{-2}) \biggr)d\phi^a d\phi^b\,,\label{Beig-Schmidt-spatial}
\end{align}
where $h_{ab}$ is the Lorentzian hyperboloid metric,
\begin{align}
h_{ab}d\phi^a d\phi^b& = -d\tau^2 + \cosh^2\!\tau \gamma_{AB}dx^Adx^B\, ,
\end{align}
and $\sigma, k_{ab}, i_{ab}$, and $j_{ab}$ are fields on the hyperboloid. The field $k_{ab}$ is taken to be trace-free,
\begin{align}\label{tracek0}
k_{ab}h^{ab} = 0\,,
\end{align}
and of the form
\begin{align}
\label{kfromPhi0}
k_{ab} =2 (D_aD_b + h_{ab})\psi\,,
\end{align}
where, $D_{a}$ denote three dimensional covariant derivative with respect to the metric $h_{ab}$. 
The assumption \eqref{kfromPhi0} for the form of $k_{ab}$ is equivalent to the vanishing of the magnetic part of the Weyl tensor at first order in the Beig-Schmidt expansion \cite{Ashtekar:1978zz, BeigSchmidt}. The choice for $k_{ab}$ in eq.~\eqref{kfromPhi0} introduces the field $\psi$, which plays a key role in the following. Eq.~\eqref{tracek0} implies that $\psi$ satisfies
\begin{align}\label{Phieqn0}
(D^2+3)\psi=0\,.
\end{align}
Under the supertranslation generated by the function $\omega$, $k_{ab}$ transforms as,
\be
\delta_\omega k_{ab} = 2 (D_aD_b + h_{ab})\omega\,,
\ee
which in turn implies, 
\be
\delta_\omega \psi = \omega\,.
\ee
The above results and notations from \cite{Compere:2023qoa, Chakraborty:2021sbc} will be followed throughout the rest of this work in connection with spacelike infinity.

With this notation, the supertranslation invariant Lorentz charge for the hyperboloid Killing vector $\xi$ at spacelike infinity is defined as, 
\bea \label{24X24.01}
Q_{\xi}^{i^{0}(\textrm{inv})}=Q_{\xi}^{i^{0}(\textrm{L})}-Q_{\omega=\mathcal{L}_{\xi}\psi}^{i^{0}(\textrm{ST})}~,
\eea
where $\mathcal{L}_{\xi}$ denotes the Lie derivative, and $Q_{\xi}^{i^{0}(\textrm{L})}$ is the Lorentz charge, whose explicit expression is given by \cite{Compere:2023qoa},
\bea
Q_{\xi}^{i^{0}(\textrm{L})} &= & -\frac{1}{8\pi}\int_{S^{2}}d^{2}x\sqrt{-h}\,\xi^{b}n^{a}
\Bigg[-j_{ab}+\frac{1}{2}i_{ab}+\frac{1}{2}k_{a}^{~c}k_{cb} 
\nonumber 
\\ 
&& 
\qquad \qquad \qquad 
+h_{ab}\left(8\sigma^{2}+D_{c}\sigma D^{c}\sigma-\frac{1}{8}k_{cd}k^{cd}+k_{cd}D^{c}D^{d}\sigma\right)\Bigg]~,
\eea
and $Q_{\omega}^{i^{0}(\textrm{ST})}$ is the the supertranslation charge, which reads,
\bea
Q_{\omega}^{i^{0}(\textrm{ST})}=\frac{1}{4\pi}\int_{S^{2}}d^{2}x\sqrt{-h}\,n^{a}\left(\omega D_{a}\sigma-\sigma D_{a}\omega\right)~.
\eea
Here, $n^{a}$ is the future-directed unit normal to the $S^{2}$ cross-section of the Lorentzian hyperboloid. 

Note that $\omega$ is assumed to be an independent quantity, and hence, we have the following commutation relation,
\bea
\left\{Q_{\omega}^{i^{0}(\textrm{ST})},Q_{\xi}^{i^{0}(\textrm{L})}\right\}=-Q_{\mathcal{L}_{\xi}\omega}^{i^{0}(\textrm{ST})}~,
\eea
which shows that the Lorentz charges are not invariant under supertranslations, as expected. While, for the invariant charge $Q_{\xi}^{i^{0}(\textrm{inv})}$ defined above, it follows that, 
\bea
\left\{Q_{\omega}^{i^{0}(\textrm{ST})},Q_{\xi}^{i^{0}(\textrm{inv})}\right\}&=&\left\{Q_{\omega}^{i^{0}(\textrm{ST})},Q_{\xi}^{i^{0}(\textrm{L})}\right\}-\left\{Q_{\omega}^{i^{0}(\textrm{ST})},Q_{\mathcal{L}_{\xi}\psi}^{i^{0}(\textrm{ST})}\right\}
\nn
\\
&=&-Q_{\mathcal{L}_{\xi}\omega}^{i^{0}(\textrm{ST})}+\delta_{\omega}Q_{\mathcal{L}_{\xi}\psi}^{i^{0}(\textrm{ST})}~.
\eea
Let us look at the second term in the above expression in more detail. We have
\bea
\delta_{\omega}Q_{\mathcal{L}_{\xi}\psi}^{i^{0}(\textrm{ST})}&=&\delta_{\omega}\left[\frac{1}{4\pi}\int_{S^{2}}d^{2}x\sqrt{-h}\,n^{a}\left\{\xi^{b}D_{b}\psi D_{a}\sigma-\sigma D_{a}\left(\xi^{b}D_{b}\psi\right)\right\} \right]
\nn
\\
&=&\frac{1}{4\pi}\int_{S^{2}}d^{2}x\sqrt{-h}\,n^{a}\left\{\xi^{b}\left(D_{b}\delta_{\omega}\psi \right)D_{a}\sigma-\sigma D_{a}\left(\xi^{b}D_{b}\delta_{\omega}\psi\right)\right\}
\nn
\\
&=&\frac{1}{4\pi}\int_{S^{2}}d^{2}x\sqrt{-h}\,n^{a}\left\{\xi^{b}D_{b}\omega D_{a}\sigma-\sigma D_{a}\left(\xi^{b}D_{b}\omega\right)\right\}
\nn
\\
&=&Q_{\mathcal{L}_{\xi}\omega}^{i^{0}(\textrm{ST})}
\eea
where, we have used the property $\delta_{\omega}\psi=\omega$, and hence it follows that, 
\bea \label{24X24.03}
\left\{Q_{\omega}^{i^{0}(\textrm{ST})},Q_{\xi}^{i^{0}(\textrm{inv})}\right\}=0,
\eea
implying that the invariant Lorentz charges are indeed invariant under supertranslations. 

A similar definition for the invariant Lorentz charges can be given at timelike infinities $i^{\pm}$. Let us introduce the relevant notation.  Asymptotically flat spacetimes at timelike infinity are defined as a class of  spacetimes that admits the following expansion \cite{Chakraborty:2021sbc, Compere:2023qoa}, 
\begin{align}
    ds^2 & = \left(-1 - \frac{2\sigma}{\tau}-\frac{\sigma^2}{\tau^2}+ o(\tau^{-2})\right)d\tau^2 + o(\tau^{-2})\tau d\tau d\phi^a \nonumber \\
    &+\tau^2\biggr(h_{ab}+\tau^{-1}(k_{ab}-2\sigma h_{ab}) + \frac{\log \tau}{\tau^2}i_{ab} + \tau^{-2}j_{ab} + o(\tau^{-2})\biggr)d\phi^ad\phi^b\,, \label{Beig-Schmidt-Timelike}
\end{align}
where $h_{ab}$ now is the metric on the unit EAdS$_3$ spacelike hyperboloid, which can be expressed as, 
\begin{align}\label{EAdS}
h_{ab}d\phi^{a}d\phi^{b}=d\rho^{2}+\sinh^{2}\rho \gamma_{AB}dx^{A}dx^{B}~,
\end{align}
and $\sigma$, $k_{ab}$, $i_{ab}$, and $j_{ab}$ are fields on $i^{\pm}$.
Under the supertranslation generated $\omega$, the field $k_{ab}$ transforms as
\be
\delta_\omega k_{ab} =  -2 (D_a D_b - h_{ab})\omega\,.
\label{delta_omega_T}
\ee
Additionally, as in the case of the spacelike infinity  discussed above, the field $k_{ab}$ can be chosen to be of the form,
\begin{align}
\label{kfromPhi}
k_{ab} = -2(D_aD_b - h_{ab})\psi\,,
\end{align}
so that, like the situation at spacelike infinity, here also,
\be
\delta_\omega \psi = \omega\,.
\ee
With this notation, the supertranslation invariant Lorentz charge at timelike infinity is defined to be,
\bea \label{24X24.02}
Q_{\xi}^{i^{\pm}(\textrm{inv})}=Q_{\xi}^{i^{\pm}(\textrm{L})}-Q_{\omega=\mathcal{L}_{\xi}\psi}^{i^{\pm}(\textrm{ST})}~,
\label{inv-i-plus}
\eea
where in the notation of \cite{Compere:2023qoa} the Lorentz charge takes the following form,
\bea \label{24VII24.01}
Q_{\xi}^{i^{\pm}(\textrm{L})} &=& \pm\frac{1}{8\pi}\int_{S}d^{2}x\sqrt{q}\,\xi^{b}r^{a}
\Bigg[-j_{ab}+\frac{1}{2}i_{ab}+\frac{1}{2}k_{a}^{~c}k_{cb} \nonumber \\ & & \qquad \qquad +h_{ab}\left(8\sigma^{2}-D_{c}\sigma D^{c}\sigma-\frac{1}{8}k_{cd}k^{cd}-k_{cd}D^{c}D^{d}\sigma\right)\Bigg]~, 
\eea
and the supertranslation charge reads,
\bea
Q_{\omega}^{i^{\pm}(\textrm{ST})}=\frac{1}{4\pi}\int_{S}d^{2}x\sqrt{q}\,r^{a}\left(\omega D_{a}\sigma-\sigma D_{a}\omega\right)~.
\eea
Here, $S$ is a closed two-surface embedded in the spacelike hyperboloid $i^{+}$ or $i^{-}$ with $r^{a}$ being the outward pointing normal vector on $S$. Furthermore, $q$ is the determinant of the induced metric on the closed two-surface $S$, and $\xi^{a}$ is a Killing vector of the three-dimensional hyperboloid. The commutator between the invariant charge and the supertranslation charge is seen to vanish,
\bea \label{24X24.04}
\left\{Q_{\omega}^{i^{\pm}(\textrm{ST})},Q_{\xi}^{i^{\pm}(\textrm{inv})}\right\}=0\,,
\eea
confirming that $Q_{\xi}^{i^{\pm}(\textrm{inv})}$ is indeed a supertranslation invariant Lorentz charge. 

To the best of our knowledge, the proposed definitions for supertranslation invariant Lorentz charges at spacelike/timelike infinity are novel. In particular, the vanishing commutators as presented in eqs.~\eqref{24X24.03}, \eqref{24X24.04} are a non-trivial check confirming that our definition of Lorentz charges is indeed supertranslation invariant. Note that the definition of supertranslation invariant Lorentz charges at null infinity is available in \cite{Compere:2023qoa} (see eq. E.5, in particular) and shares similarity with our definition.

\section{Connecting invariant charges at null and spacelike/timelike infinity}
\label{sec:matching_CWY}

In the previous section, we proposed definitions for supertranslation invariant Lorentz charge at spacelike and timelike infinities. Now, we wish to understand how these charges behave when these hyperboloids reach null infinity.  

We start by examining the invariant Lorentz charges defined at future null infinity $\mathcal{J}^{+}$. A definition was given in \cite[eq.~(23)]{Fuentealba:2023syb} 
\bea \label{01II24.02}
Q_{Y,~\textrm{FHT}}^{\mathcal{J}^{\pm}\textrm{(inv)}}&=&Q_{Y}^{\rm{old}}-4\int_{S^{2}} d\Omega\left[Y^{A}\partial_{A}C -\frac{1}{2}C\left(\nabla_{A} Y^{A}\right)\right]m,
\eea
where $Y^{A}$ is a conformal Killing vector field on the round two-sphere, $\nabla_{A}$ is the covariant derivative compatible with the round metric on the two-sphere, and 
\bea
Q_{Y}^{\rm{old}}=2\int_{S^{2}} d\Omega\,Y^{A}N_{A}\,.
\eea
In writing these expressions, we are assuming familiarity with the Bondi-Sachs formalism, and we are following the notation of \cite{Compere:2023qoa}. The function $m$ is the mass aspect, $N^A$ is the angular momentum aspect, and $d\Omega = \sin \theta d \theta d\phi$ is the two-dimensional solid angle. The field 
$C$ is the electric (parity-even) part of the shear tensor $C_{AB}$,
\begin{align}\label{Cdecomp}
C_{AB} = (-2\nabla_A\nabla_B + \gamma_{AB}\nabla^2)C + \epsilon_{D(A}\nabla_{B)}\nabla^{D}\Psi~,
\end{align}
while $\Psi$ is the magnetic (parity-odd) part.

Ref.~\cite[eq.~(E.6)]{Compere:2023qoa} also defines a supertranslation invariant Lorentz charge at null infinity. It is slightly more convenient to work with that definition. They define,
\be 
Q_{Y,~\textrm{CGW}}^{\mathcal{J}^{\pm}\textrm{(inv)}}=\frac{1}{8\pi}\int_{S^{2}} d\Omega Y^{A}\left(N_{A} -3 m\partial_{A} C-C \partial_{A}m\right). \label{CGW-inv}
\ee
Relation to $Q_{Y,~\textrm{FHT}}^{\mathcal{J}^{\pm}\textrm{(inv)}}$ can be seen as follows: we first rewrite the charge defined in eq.~\eqref{CGW-inv} as,
\bea 
Q_{Y,~\textrm{CGW}}^{\mathcal{J}^{\pm}\textrm{(inv)}} = \frac{1}{16\pi} Q_{Y}^{\rm{old}}
-\frac{1}{8\pi}\int_{S^{2}}d\Omega\,\bigg[2 m Y^{A}\partial_{A} C-m\nabla_{A}Y^{A}\bigg]
-\frac{1}{8\pi}\int_{S^{2}}d\Omega\,\nabla_{A}\left(mCY^{A}\right),
\label{10XI23.01}
\eea
where in eq.~\eqref{10XI23.01} we have rewritten the term $Y^{A}C\partial_{A}m$ 
as a total derivative term along with additional contributions. The total derivative vanishes. Hence, the CGW charge integral in eq.~\eqref{CGW-inv} becomes,
\bea \label{01II24.01}
Q_{Y,~\textrm{CGW}}^{\mathcal{J}^{\pm}\textrm{(inv)}}&=&\frac{1}{8\pi}\int_{S^{2}} d\Omega\,Y^{A} N_{A}-\frac{1}{4\pi}\int_{S^{2}}d\Omega\left[Y^{A}\partial_{A} C -\frac{1}{2} C\left(\nabla_{A}Y^{A}\right)\right]m~.
\eea
This expression is the same as eq.~\eqref{01II24.02} up to an overall numerical factor,
$Q_{Y,~\textrm{FHT}}^{\mathcal{J}^{\pm}\textrm{(inv)}}=16\pi \,Q_{Y,~\textrm{CGW}}^{\mathcal{J}^{\pm}\textrm{(inv)}}.$ We note that the equality between FHT and CGW charges is mentioned in \cite{Fuentealba:2023syb}. These charges are defined on the cross-section of the null infinity in non-radiative regions. They differ by boundary term, and the vanishing boundary contribution of the compact sphere plays an important role in the equality of these charges. 

From now onwards we drop the subscripts FHT and CGW  and work exclusively with the CGW expressions.

The future boundary of the future null infinity is reached by taking $u\to \infty$. In this limit,
\bea 
Q_{Y}^{\mathcal{J}^{+}_{+}(\textrm{inv})} :=\lim_{u \to \infty}Q_{Y}^{\mathcal{J}^{+}(\textrm{inv})}=\lim_{u \to \infty} \frac{1}{8\pi} \int_{S^{2}} d\Omega \bigg[Y^{A}(N_{A}-2m\partial_{A}C)+m C\nabla_{A}Y^{A}\bigg]~.
\eea
In order to determine the above limit, we consider the asymptotic $u$-expansion of quantities appearing in the above expression (see eqs. (126-128) of \cite{Compere:2023qoa}),
\bea 
m &=& m^{(0)}+m^{(1)} u^{-1}+o(u^{-1}),
\\
N_{A}&=& N_{A}^{\log} \log u+N_{A}^{(0)}+o(1),
\\
C_{AB}&=&C_{AB}^{(0)}+C_{AB}^{(1)} u^{-1}+o(u^{-1}).
\eea
In the $u \to \infty$ limit, $\log u$ can lead to divergences.  Using Bondi expansion and the asymptotic equations of motion, we note that $N_{A}^{\log}$ can be expressed as (see eq.(130) of \cite{Compere:2023qoa}),
\bea \label{NAlog}
N_{A}^{\log}=\partial_{A}m^{(1)}+\frac{1}{4} \epsilon_{AB} \nabla^{B}\nabla^{2}(\nabla^{2}+2) \Psi^{(1)}~,
\eea
where $\nabla^{2}\equiv \nabla_{B}\nabla^{B}$ and $\Psi^{(1)}$ is magnetic part of $C_{AB}^{(1)}$, defined analogously in equation \eqref{Cdecomp}. Thus, we obtain the following expression,
\bea 
Q_{Y}^{\mathcal{J}^{+}_{+}(\textrm{inv})}&=& \lim_{u \to \infty}\frac{\log u}{8\pi} \int_{S^{2}} d\Omega ~Y^{A} \bigg(\partial_{A}m^{(1)}+\frac{1}{4} \epsilon_{AB} \nabla^{B}\nabla^{2}(\nabla^{2}+2) \Psi^{(1)}\bigg)
\nn
\\
&& + \frac{1}{8\pi} \int_{S^{2}} d\Omega \bigg[Y^{A}(N_{A}^{(0)}-2m^{(0)}\partial_{A}C^{(0)})+m^{(0)}C^{(0)}\nabla_{A}Y^{A}\bigg]~.
\eea 
The coefficient of the $\log u$ term in the above expression can be written by an integration-by-parts as
\bea \label{21VI24.01}
\textrm{coeff.~of}\, \log u=\frac{1}{8\pi} \int_{S^{2}} d\Omega \bigg[-m^{(1)}\nabla_{A}Y^{A}+\frac{1}{2} \epsilon_{AB} (\nabla^{A}Y^{B})\nabla^{2}(\nabla^{2}+2) \Psi^{(1)}\bigg]~.
\eea
To simplify this expression further, we note that a conformal Killing vector field $Y^{A}$ on the sphere can be written as, 
\bea\label{confkilling}
Y_{A}=\nabla_{A} G+\epsilon_{AB}\nabla^{B} H~,
\eea
where $G$ and $H$ are two functions satisfying
\be
(\nabla^{2}+2)\nabla^{2} G= (\nabla^{2}+2)\nabla^{2} H = 0\,. 
\label{G-H-funcs}
\ee
Inserting this decomposition in eq.~\eqref{21VI24.01} yields,
\bea 
\textrm{coeff.~of}\, \log u=\frac{1}{8\pi} \int_{S^{2}} d\Omega \bigg[-m^{(1)}\nabla^{2} G+\frac{1}{2}\epsilon_{AB}\nabla^{A}(\epsilon^{BC}\nabla_{C}H)\nabla^{2}(\nabla^{2}+2)\Psi^{(1)}\bigg]~.
\eea
Now, using the asymptotic equation of motion (\textcolor{blue}{see eq. (129) of \cite{Compere:2023qoa}})
\be
m^{(1)}=-\frac{1}{4} \nabla^{2}(\nabla^{2}+2)C^{(1)}, \label{m1-eq}
\ee
we obtain
\bea 
\textrm{coeff.~of}\, \log u=\frac{1}{8\pi}\int_{S^{2}} d \Omega \bigg[\frac{1}{4}(\nabla^{2}G)\nabla^{2}(\nabla^{2}+2)C^{(1)}-\frac{1}{2}\nabla^{2}H \nabla^{2}(\nabla^{2}+2)\Psi^{(1)}\bigg]~.
\eea
To simplify further, we use the identity 
\be
\nabla^{2}(\nabla^{2}+2)f= (\nabla^{2}+2)\nabla^{2}f\,,
\ee 
for an arbitrary scalar function $f$. Upon integration by parts, the coefficient of $\log u$ becomes,
\bea 
\textrm{coeff.~of}\, \log u=\frac{1}{{8\pi}} \int_{S^{2}} d\Omega \bigg[\frac{1}{4}\left\{(\nabla^{2}+2)\nabla^{2} G\right\} \nabla^{2} C^{(1)}-\frac{1}{2}\left\{(\nabla^{2}+2)\nabla^{2} H\right\} \nabla^{2} \Psi^{(1)} \bigg]~.
\eea
From \eqref{G-H-funcs}, we conclude that the coefficient of the $\log u$ term vanishes. Thus, the $\log u$ term does not feature in the asymptotic analysis at null infinity, which was also the case in \cite{Compere:2023qoa}. The invariant Lorentz charge at the future boundary of the future null infinity then becomes,
\bea
Q_{Y}^{\mathcal{J}^{+}_{+}(\textrm{inv})}=\frac{1}{8\pi} \int_{S^{2}} d\Omega \bigg[Y^{A}\left(N_{A}^{(0)}-2m^{(0)}\partial_{A}C^{(0)}\right)+m^{(0)}C^{(0)}\nabla_{A}Y^{A}\bigg]~.
\eea 

On the other hand, at timelike infinity $(i^{+})$, the invariant charge, as proposed in this work, takes the form \eqref{inv-i-plus}.
In order to relate this expression to the invariant charge at the future end of the future null infinity, we consider the two-surface $S$ to be $\rho=\textrm{constant}$ and take the limit $\rho \to \infty$, cf.~eq.~\eqref{EAdS}. For this choice, $\sqrt{q}=\sinh^{2}\rho$ and $r^{a}=\delta^{a}_{\rho}$, and $\xi^{b}=Y^{A}e^{b}_{A}-(1/2)(\nabla_{C}Y^{C})e^{b}_{\rho}$, where $Y^{A}$ is the conformal Killing vector field on the two-sphere~\cite{Compere:2023qoa}. Here, we have introduced the projection operator $e^{b}_{A}=(\partial \phi^{b}/\partial x^{A})$, and $e^{b}_{\rho}=(\partial \phi^{b}/\partial \rho)=\delta^{b}_{\rho}$. In $\rho\to \infty$ limit the field $\psi$ matches with the field $C$ as 
\be
\lim_{\rho\to \infty} \psi (\rho, x^A) = (1/2)e^{\rho}C^{(0)}.
\ee 
In this limit, the part corresponding to the supertranslation charge reads,
\bea
Q_{\omega=\mathcal{L}_{\xi}\psi}^{i^{+}_{\partial}(\textrm{ST})}&=&\lim_{\rho\to \infty}Q_{\omega=\mathcal{L}_{\xi}\psi}^{i^{+}(\textrm{ST})}
\nn
\\
&=&\lim_{\rho\to \infty}\frac{1}{4\pi}\int_{S^{2}}d\Omega\frac{e^{2\rho}}{4}\,\Bigg[\left\{Y^{A}\partial_{A}\psi-\frac{1}{2}\left(\nabla_{A}Y^{A}\right)\partial_{\rho}\psi\right\}\partial_{\rho}\sigma \nn \\
&\,&\qquad -\sigma \partial_{\rho}\left\{Y^{A}\partial_{A}\psi-\frac{1}{2}\left(\nabla_{A}Y^{A}\right)\partial_{\rho}\psi\right\}\Bigg]
\nn
\\
&=&\lim_{\rho\to \infty}\frac{1}{4\pi}\int_{S^{2}}d\Omega\frac{e^{2\rho}}{4}\,\Bigg[\left\{\frac{1}{2}e^{\rho}Y^{A}\partial_{A}C^{(0)}-\frac{1}{4}\left(\nabla_{A}Y^{A}\right)e^{\rho}C^{(0)}\right\}6m^{(0)}e^{-3\rho}
\nn
\\
&\,&\qquad +2m^{(0)}e^{-3\rho} \partial_{\rho}\left\{\frac{1}{2}e^{\rho}Y^{A}\partial_{A}C^{(0)}-\frac{1}{4}\left(\nabla_{A}Y^{A}\right)e^{\rho}C^{(0)}\right\}\Bigg]
\nn
\\
&=&\frac{1}{4\pi}\int_{S^{2}}d\Omega\,m^{(0)}\Bigg[Y^{A}\partial_{A}C^{(0)}-\frac{1}{2}\left(\nabla_{A}Y^{A}\right)C^{(0)}\Bigg]~,
\eea
where in the third line, we have used the result that $\sigma=-2m^{(0)}e^{-3\rho}$ in the asymptotic limit.

Similarly, in the  $\rho \to \infty$ limit Lorentz charge \eqref{24VII24.01}  reads (with $h_{\rho A}=0$, $h_{\rho \rho}=1$),
\bea \label{Qtimelikeinv}
Q_{\xi}^{i^{+}_{\partial}(\textrm{L})}&=&\lim_{\rho\to \infty}Q_{\xi}^{i^{+}(\textrm{L})}
\nn
\\ 
&=&\frac{1}{8\pi}\int_{S^{2}}d\Omega\,\frac{e^{2\rho}}{4}Y^{A}
\Big[-j_{\rho A}+\frac{1}{2}i_{\rho A}+\frac{1}{2}k_{\rho}^{~c}k_{cA}\Big]
-\frac{1}{8\pi}\int_{S^{2}}d\Omega\,\frac{e^{2\rho}}{8}\left(\nabla_{C}Y^{C}\right)
\Big[-j_{\rho \rho}
\nn
\\
&\,&\qquad +\frac{1}{2}i_{\rho \rho}+\frac{1}{2}k_{\rho}^{~c}k_{c\rho}+\big(8\sigma^{2}-D_{c}\sigma D^{c}\sigma-\frac{1}{8}k_{cd}k^{cd}-k_{cd}D^{c}D^{d}\sigma\big)\Big]~.
\eea
The above charge neatly separates into two parts: we call the part multiplying $Y^A$ to be  $Q^{(1)}$ and the part multiplying $\nabla_{C}Y^{C}$ to be $Q^{(2)}$.  $Q^{(1)}$ depends on $j_{\rho A}$, $i_{\rho A}$ and $k_{\rho}^{c}k_{cA}$. In the   $\rho \to \infty$ limit, these quantities behave as\footnote{Equations \eqref{24VII24.02}, \eqref{24VII24.03}, \eqref{24VII24.04} are obtained from equations (157), (155), (144-146) of \cite{Compere:2023qoa}, respectively.}
\bea 
j_{\rho A}&=&-\nabla^{B}C_{AB}^{(1)}+e^{-2\rho} \Big[-4N_{A}^{(0)}-C_{AC}^{(0)}\nabla_{B}C^{(0)BC} \nn \\ 
&& -12\partial_{A}m^{(1)} +(4\rho-2)N_{A}^{\log}-\nabla^{B}C_{AB}^{(1)}\Big]+o(e^{-3\rho}), \label{24VII24.02}\\
i_{\rho A}&=&-4e^{-2\rho}N_{A}^{\log}+o(e^{-3\rho})\,,
\label{24VII24.03} \\ 
k_{\rho}^{~c}k_{cA} &=& k_{\rho \rho}k^{\rho}_{A}+k_{\rho B}k^{B}_{A}
=
{-}4e^{-2\rho}C^{(0)B}_{A}\nabla^{C}C^{(0)}_{CB}+o(e^{-2\rho})~. \label{24VII24.04}
\eea 
Thus, for $Q^{(1)}$ we obtain the expression
\bea 
Q^{(1)}=\frac{1}{8\pi}\int_{S^{2}}d\Omega\,Y^{A}
\Big[\frac{1}{4}\left(1+e^{2\rho}\right)\nabla^{B}C^{(1)}_{AB}+N_{A}^{(0)}+3\partial_{A}m^{(1)}-\frac{1}{4}C^{(0)}_{AC}\nabla_{B}C^{(0)BC}-\rho N_{A}^{\log}\Big]~.
\eea
Similarly, for
$Q^{(2)}$ in the $\rho\to \infty$ limit, we need
\bea 
j_{\rho \rho}&=&16m^{(1)}e^{-2\rho}+o(e^{{-2\rho}})~,
\\
k_{cd}k^{cd}&=&4e^{-2\rho}C^{(0)}_{AB}C^{(0)AB}+o(e^{{-2\rho}})~,
\eea
and $i_{\rho \rho}=o(e^{-2\rho})$, $k_{\rho c}k^{c}_{\rho}={o}(e^{-3\rho})$, $\sigma^{2}={o}(e^{-5\rho})$, $k_{cd}D^{c}D^{d}\sigma= o(e^{-5\rho})$, $ D_{c}\sigma D^{c}\sigma = o(e^{-5\rho})$.\footnote{These relations can be obtained from equations (138), (144-146), (154-157) of \cite{Compere:2023qoa}.} As a result, 
\bea 
Q^{(2)}&=&\frac{1}{8\pi}\int_{S^{2}}d\Omega\,\left(\nabla_{A}Y^{A}\right)
\Big[2m^{(1)}+\frac{1}{16}C^{(0)}_{AB}C^{(0)AB}\Big]~.
\eea
Thus the asymptotic Lorentz charge at $i^{+}$ becomes,
\bea 
Q_{\xi}^{i^{+}_{\partial}(\textrm{L})}&=&\frac{1}{8\pi}\int_{S^{2}}d\Omega\,Y^{A}N_{A}^{(0)} \nonumber \\ 
&+&\frac{1}{8\pi}\int_{S^{2}}d\Omega\,Y^{A}\Big[3\partial_{A}m^{(1)}-\frac{1}{4}C^{(0)}_{AC}\nabla_{B}C^{(0)BC}-\rho N_{A}^{\log}+\frac{1}{4}\left(1+e^{2\rho}\right)\nabla^{B}C^{(1)}_{AB}\Big]
\nn
\\
&+&\frac{1}{8\pi}\int_{S^{2}}d\Omega\,\left(\nabla_{A}Y^{A}\right)
\Big[2m^{(1)}+\frac{1}{16}C^{(0)}_{AB}C^{(0)AB}\Big]
\nn
\\
&=&\frac{1}{8\pi}\int_{S^{2}}d\Omega\,Y^{A}N_{A}^{(0)}\,.
\label{final}
\eea
In arriving at the last line we have used several results. First, we note that, 
\begin{align} 
-\frac{1}{32\pi}&\int_{S^{2}}d\Omega\,Y^{A}C^{(0)}_{AC}\nabla_{B}C^{(0)BC}
+\frac{1}{32\pi}\frac{1}{4}\int_{S^{2}}d\Omega\,\left(\nabla_{A}Y^{A}\right)C^{(0)}_{AB}C^{(0)AB}
\nn
\\
&=\textrm{total~derivative}-\frac{1}{32\pi}\int_{S^{2}}d\Omega\,Y^{A}\left[C^{(0)}_{AC}\nabla_{B}C^{(0)BC}+\frac{1}{4}\nabla_{A}\left(C^{(0)}_{AB}C^{(0)AB}\right) \right]
\nn
\\
&=\textrm{total~derivative}\,,
\end{align}
where we have used \cite[eq.~(110)]{Compere:2023qoa}. Next, using asymptotic equations of motion, we can write $m^{(1)}$ as \eqref{m1-eq}, which along with conformal Killing condition, eq.~\eqref{confkilling} and eq.~\eqref{G-H-funcs} can be used to show that all the $m^{(1)}$ dependent terms contribute only total derivatives. Similarly, the term with $N_{A}^{\log}$ and $\nabla^{B}C_{AB}^{(1)}$ also only contributes total derivatives. Since the boundary is compact, it follows that these total derivative terms vanish identically. 


Therefore,
\bea 
Q_{\xi}^{i^{+}_{\partial}(\textrm{inv})}=\frac{1}{8\pi}\int d\Omega \bigg[Y^{A}(N_{A}^{(0)}-2m^{(0)}\partial_{A}C^{(0)})+m^{(0)}C^{(0)}\partial_{A}Y^{A}\bigg]=Q_{Y}^{\mathcal{J}^{+}_{+}(\textrm{inv})}~.
\eea
Thus, the supertranslation invariant Lorentz charge defined at null infinity in the $u\to \infty$ limit is the same as the invariant Lorentz charge proposed in this work at timelike infinity. Similarly, on the one hand, taking the limit $u\to -\infty $ of the invariant Lorentz charge defined at future null infinity, we obtain the invariant charge at $\mathcal{J}^{+}_{-}$. On the other hand, evaluating the invariant charge at spacelike infinity $i^{0}$ in the $\tau\to \infty$ limit gives the same expression as at $\mathcal{J}^{+}_{-}$. A similar equality holds for $\mathcal{J}^{-}_{-}$ and $i_-$ and also for $\mathcal{J}^{-}_{+}$ and $\tau\to - \infty$ limit of $i_0$.

\section{Comparing with Chen-Wang-Yau angular momentum at null infinity}
\label{sec:CWY_and_others}

In this section, we compare the invariant charge defined by ref.~\cite{Compere:2023qoa}, cf.~\eqref{CGW-inv}, with the invariant charge defined by Chen, Wang, and Yau (CWY) \cite{Chen:2021szm, Chen:2022fbu}. The CWY definition of supertranslation invariant quasi-local angular momentum at null infinity is
\bea \label{2II24.03}
Q^{\mathcal{J}^{\pm}\,\textrm{(inv)}}_{Y\,\textrm{(CWY)}}=Q^{\textrm{DS}}+\frac{1}{8\pi} \int_{S^{2}} d\Omega ~m Y^{A} \nabla_{A}\mathcal{C}\,,
\eea
where $Q^{\textrm{DS}}$ is the Dray-Streubel angular momentum \cite{Dray:1984rfa}, 
\bea\label{QDS}
Q^{\textrm{DS}}=\frac{1}{8\pi} \oint_{S^{2}} d\Omega~ Y^{A}\bigg(N_{A}-\frac{1}{4} C_{AB} \nabla_{D} C^{DB}\bigg)\,,
\eea
and where $\mathcal{C}$ is the unique solution to the equation 
\bea \label{CWYC}
\nabla^{2}\left(\nabla^{2}+2\right)\mathcal{C}=2\nabla^{A}\nabla^{B}C_{AB}\,.
\eea 
Expression \eqref{2II24.03} can be taken to be a well-defined notion of a ``pure rotation'' as opposed to a ``rotation plus supertranslation'' on a given cross-section~\cite{Chen:2022fbu}. In this definition, the vector field $Y^{A}$ is a Killing vector on the two-sphere (as opposed to a conformal Killing vector).

Our first observation is that the scalar $\mathcal{C}$ is directly related to the trace of the Bondi shear $C_{AB}$. 
Taking double divergence of \eqref{Cdecomp}, we get 
\bea  \label{2II24.01}
\nabla^{A}\nabla^{B}C_{AB}=-\nabla^{2}\left(\nabla^{2}+2\right)C.
\eea
Comparing  eq.~\eqref{CWYC} and eq.~\eqref{2II24.01} we obtain the desired relation $\mathcal{C} = - 2C$.

To proceed further, \emph{we assume that the magnetic contribution to the shear tensor vanishes}. Therefore, the shear tensor $C_{AB}$  can be written as cf.~\eqref{Cdecomp},
\bea \label{2II24.02}
C_{AB}=\left(-2\nabla_{A} \nabla_{B}+\gamma_{AB} \nabla^{2}\right)C\,.
\eea
Taking the divergence of eq.~\eqref{2II24.02}, we obtain, 
\bea 
\nabla_{A}C^{AB}=-\nabla^{B}\left(\nabla^{2}+2\right)C\,.
\eea
After careful manipulation, using the above identity and the result that $\mathcal{C}=-2C$, the second term in eq.~\eqref{QDS} can be written as 
\bea  \nn
Y^{A} C_{AB} \nabla_{D} C^{DB} &=& - \nabla^{B} \bigg[Y^{A} C_{AB} (\nabla^{2}+2)C\bigg]-\frac{1}{2} \nabla^{A}\bigg[Y^{A} \left\{\left(\nabla^{2}+2\right)C\right\}^{2}\bigg]
\\
&& +\left(\nabla^{A}Y^{B}\right)C_{AB}\left(\nabla^{2}+2\right)C+\frac{1}{2}\left(\nabla^{A}Y_{A}\right) \left[\left(\nabla^{2}+2\right)C\right]^{2}\,.
\eea
Upon integration over a two-sphere, the first two terms in the above expression vanish. Under the assumption that $Y^A$ is a Killing vector on the sphere, the remaining two terms also vanish. Therefore, the supertranslation invariant angular momentum charge defined in \cite{Chen:2021szm, Chen:2022fbu} becomes,
\bea \nn
Q^{\mathcal{J}^{\pm}\,\textrm{(inv)}}_{Y\,\textrm{(CWY)}}=
&=&\frac{1}{8\pi} \int_{S^{2}} d\Omega~ Y^{A}\bigg(N_{A}-\frac{1}{4} C_{AB} \nabla_{D} C^{DB}\bigg)-\frac{1}{4\pi}\int_{S^{2}} d\Omega ~m Y^{A} \partial_{A} C 
\nn 
\\ 
&=&
\frac{1}{8\pi} \int_{S^{2}} d\Omega~ Y^{A}\bigg(N_{A}-2m\partial_{A}C\bigg)\,.
\label{2II24.05}
\eea
For a Killing vector field $Y^{A}$ this expression matches with eq.~\eqref{01II24.01}.

In summary, we have shown the equivalence of the CWY charge to the CGW charge at null infinity. Since we have already shown the equivalence between CGW and FHT  supertranslation invariant Lorentz charges, it follows that all these definitions are equivalent at null infinity. Further, the charges defined in this work at spacelike/timelike infinity match with the CGW charges at null infinity, and hence, there is a complete prescription for supertranslation invariant Lorentz charges at all asymptotic boundaries. A glance at the results from previous literature, e.g., \cite{Javadinezhad:2022ldc,Javadinezhad:2022hhl,Chen:2021szm,Chen:2022fbu} shows that their definitions of charges at null infinity closely follow that of CGW, and hence the above shows the one-to-one correspondence between various charges defined in the literature.

\section*{Acknowledgements}

We thank Marc Henneaux and Geoffrey Comp\`ere for discussions and email correspondence. AV additionally thanks Alok Laddha for important discussions regarding ref.~\cite{Henneaux:2023neb}. The work of AV is partly supported by SERB Core Research Grant CRG/2023/000545. The work of JH is supported in part by MSCA Fellowships CZ-UK2 $(\mbox{reg. n. CZ}.02.01.01/00/22\_010/0008115)$ from the Programme Johannes Amos Comenius co-funded by the European Union. JH also acknowledges the support from Czech Science Foundation Grant 22-14791S. Research of SC is supported by MATRICS (MTR/2023/000049) and Core Research  (CRG/2023/000934) Grants from SERB, Government of India. SC also acknowledges the warm hospitality at the Albert-Einstein Institute, Potsdam, where a part of this work was done. The visit was supported by a Max-Planck-India mobility grant.


\end{document}